\documentclass[preprint2]{aastex}

\shorttitle{the Bi-directional Moving Structures in a Coronal Bright
Point} \shortauthors{Li et al.}

\begin{document}


\title{The Bi-directional Moving Structures in a Coronal Bright Point}


\author{Dong Li\altaffilmark{1,2} \& Zongjun Ning\altaffilmark{1} \& Yingna Su\altaffilmark{1}}
\email{lidong@pmo.ac.cn \\
ningzongjun@pmo.ac.cn \\ ynsu@pmo.ac.cn}


\altaffiltext{1}{Key Laboratory of Dark Matter and Space Astronomy,
Purple Mountain Observatory, CAS, Nanjing 210008, China}
\altaffiltext{2}{Key Laboratory of Modern Astronomy and Astrophysics
(Nanjing University), Ministry of Education, Nanjing 210093, China}


\begin{abstract}
We report the bi-directional moving structures in a coronal bright
point (CBP) on 2015 July 14. It is observed by the Atmospheric
Imaging Assembly (AIA) onboard {\it Solar Dynamics Observatory
(SDO)}. This CBP has a lifetime of $\sim$10 minutes, and a curved
shape. The observations show that many bright structures are moving
intermittently outward from the CBP brightness core. Such moving
structures are clearly seen at AIA 171~{\AA}, 193~{\AA}, 211~{\AA},
131~{\AA}, 94~{\AA}, 335~{\AA} and 304~{\AA}, slit-jaw (SJI)
1330~{\AA} and 1400~{\AA}. In order to analyze these moving
structures, the CBP is cut along the moving direction with a curved
slit from the AIA and SJI images. Then we can obtain the
time-distance slices, including the intensity and
intensity-derivative diagrams, from which, the moving structures are
recognized as the oblique streaks, and they are characterized by the
bi-direction, simultaneity, symmetry, and periodicity. The average
speed is around 300~km~s$^{-1}$, while the typically period is
$\sim$90~s. All these features (including the bi-directional flows
and their periodicity) can be detected simultaneously at all the 9
wavelengths. This CBP takes place at the site between a small pair
of magnetic polarities. High time resolution observations show that
they are moving close to each other during its lifetime. These facts
support the magnetic reconnection model of the CBP and the
bi-directional moving structures could be the observational outflows
after the reconnection. Therefore, they can be as the direct
observation evidence of the magnetic reconnection.
\end{abstract}

\keywords{Sun: atmosphere; Sun: UV radiation; Sun: magnetic
reconnection}


\section{Introduction}
Bright points (BPs) are frequently detected as the small scale
brightness enhancements on solar disk. Different from the larger
scale solar burst phenomena, i.e., flares and CMEs, they are very
common phenomena in the active regions, quiet Sun or even coronal
holes. The BPs are usually observed at the solar atmosphere, i.e.,
from photosphere through chromosphere to corona. The BPs observed at
X-rays \citep{Zhang12,Zhang13,Zhang14}, extreme ultraviolet (EUV)
\citep{Ning14,Alipour15}, and ultraviolet (UV) \citep{Li12,Li13}
bands are generally called coronal bright points (CBPs). They
usually display the small roundish or loop-shaped features. The CBPs
have a typical size of about 5$-$40\arcsec, and they can last for a
few hours or longer \citep{Vaiana73,Golub74}. Statistical results
indicate that about 1500 CBPs emergence at the full solar surface
every day \citep{Golub74}. And further studies suggest that there
are 800 CBPs emergence on the entire solar surface at any given time
\citep{Zhang01}. Therefore, they are thought to be contributed to
the coronal heating. However, the number and distribution of the
CBPs on solar atmospheres are different between solar maximum and
minimum years. In other words, there are much more CBPs emergence at
sunspot minimum than that at sunspot maximum, and they appear more
uniformly during solar minimum over the solar surface. The number of
the CBPs becomes less when their lifetimes are longer
\citep{Priest94}. On the other hands, previous studies have found
that the temperature of the CBPs is about several MK, while their
electron density has an order of $\sim$10$^9$~cm$^{-3}$, which
corresponds to the emission measure (EM) of about 10$^{27}$
cm$^{-5}$ \citep{Nolte79,Tian08,Kariyappa11,Hong14,Zhang14}.

The CBPs are usually believed to be the results of the small-scale
magnetic reconnection \citep{Priest94,Zhang12,Innes13}. Observations
show that the CBPs are always associated with the small-scale
magnetic fields which have opposite-polarity poles
\citep{Golub76,Li12}. And these small magnetic fields are always new
emergence and have the fluxes of about 10$^{18}$$-$10$^{19}$ Mx
\citep{Priest94,Li13}. Previous findings further indicated that one
third of the CBPs occur at the new ephemeral magnetic regions,
whereas the other two thirds of them are lied above the cancelations
of the opposite-polarity magnetic fields \citep{Harvey85,Webb93}.
Using the observations with high time resolution, several authors
have investigated the relationship between the CBPs and the moving
behaviors of magnetic fields (or the change of the magnetic flux)
\citep[e.g.,][]{Harvey94,Alexander11,Li12,Zhang12,Chen15}. They
found that the CBP is associated with magnetic cancelation. All
these findings support the magnetic reconnection model for the CBPs.
Based on the 2-D reconnection model \citep{Sturrock64}, the
bi-directional outflows can be produced simultaneously, and the
radiation along their propagation path could be observed at EUV or
optical wavelengths. For example, the bi-directional outflows are
well detected in the transition-region explosive events
\citep{Innes97,Perez99,Innes13,Innes15} and solar flares
\citep{Su13,Liu13,Ning16}. The reconnections flows are called as the
downflows and upflows according to the moving directions (toward or
away) from the solar surface.

Similar as the quasi-periodic oscillations in the solar flare
\citep{Li15,Ning16}, some CBPs also display the quasi-periodic
oscillations, such as the periodic variation in intensity and
Doppler shifts at X-ray and EUV bands
\citep{Nolte79,Tian08b,Zhang14a,Samanta15}, or the repeated
bi-directional flows at EUV bands \citep{Ning14}. The periods of
these oscillations are ranging from seconds to tens of minutes. For
example, \cite{Ning14} detected the quasi-periods between 80 and
100~s at 9 AIA EUV (or UV) bands in a curved CBP. While
\cite{Tian08b} found the intensity oscillations with long periods
ranging from 8 to 64~min in the CBPs. Recently, \cite{Samanta15}
found the correlation between the imaging intensity and the spectral
line parameters in a CBP. All these findings suggest that the
quasi-periodic oscillations in the CBP could be explained by the
repetitive and small-scale magnetic reconnections on the solar
surface \citep{Priest94,Zhang14,Chen15,Samanta15}. And the observed
period could be the same as that of the modulated MHD waves
\citep[e.g.,][]{Chen06,Banerjee16}. This is also consistent well
with the 2-D reconnection model \citep{Sturrock64}.

Using the observations from SOHO/SUMER \citep{Wilhelm97}, the
studies of the bi-directional outflows had been breakthrough.
However, these phenomena are mostly in the short-time scale, such as
explosive events \citep[e.g.,][]{Innes97,Perez99,Innes13},
chromospheric upflow events \citep{Chae98}, and blinkers
\citep{Chae00,Brkovic04,Bewsher05}. Following the observation
results, the MHD simulations is applied to build up the models of
these events based on the magnetic reconnection
\citep[see.,][]{Innes99,Sarro99,Roussev01a,Roussev01b}. Until now,
investigation on the bi-directional outflows in the CBP is still
poorly done. In this paper, using the multi-wavelength observations
from the slit-jaw (SJI) aboard {\it Interface Region Imaging
Spectrograph (IRIS)}, the Atmospheric Imaging Assembly (AIA) and
Helioseismic and Magnetic Images (HMI) on board {\it Solar Dynamics
Observatory (SDO)}, we explore the bi-directional outflows and their
generation in a CBP at multi-wavelengths with high spatial and
temporal resolution data.

\section{Observations}
\begin{figure}
 \includegraphics[width=\columnwidth]{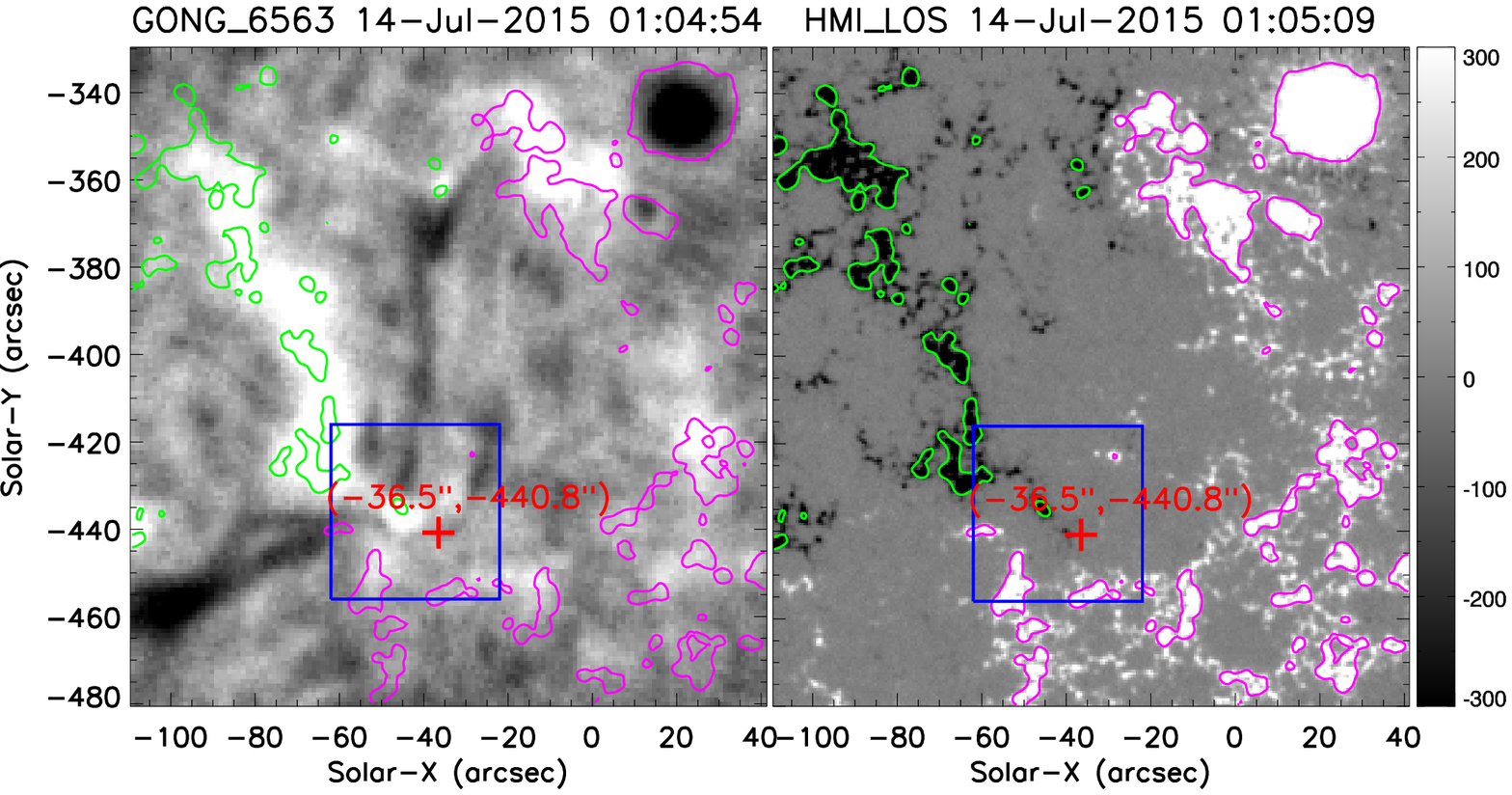}
 \caption{H$\alpha$ 6563 {\AA} image (left) from NSO/GONG,
and LOS magnetogram (right) from {\it SDO}/HMI. The contours
represent the magnetic fields at the levels of 200 (purple) and -200
(green) G, respectively. The blue box marks the region in
Fig.~\ref{cbp}, and the red plus (`+') indicates the CBP core.}
 \label{fil}
\end{figure}

The CBP studied here takes place in NOAA AR12384 on July 14 2015. It
starts at about 01:00 UT and reaches its maximum at around 01:06 UT.
Fig.~\ref{fil} (left) shows the H$\alpha$ 6563 {\AA} image observed
by NSO/GONG at 01:04:54 UT which is close to the peak time of the
CBP. There is no data between 01:05 and 01:11 UT for the NSO/GONG at
H$\alpha$ 6563 {\AA}. The contours represent the magnetic fields
from {\it SDO}/HMI \citep{Schou12}, and the levels are set at 200
(purple) and -200 (green) Gauss, respectively. The blue box marks
the region where the CBP takes place, and it is about 40\arcsec
$\times$ 40\arcsec. The CBP starts at the position of
x=$-$36.5\arcsec, y=$-$440.8\arcsec, as marked by the red plus.

{\it SDO}/AIA \citep{Lemen12} can observe the full solar disk with a
spatial scale of $\sim$0.6\arcsec\ per pixel at seven EUV and two UV
channels. The time cadence is 12 s at EUV channels, while 24 s at UV
channels. Fig.~\ref{cbp} presents the AIA images at all 9
wavelengths, such as 171~{\AA}, 193~{\AA}, 211~{\AA}, 335~{\AA},
131~{\AA}, 94~{\AA}, 304~{\AA}, 1600~{\AA} and 1700~{\AA}, and the
{\it IRIS}/slit-jaw (SJI) \citep{Dep14} images at three wavelengths,
such as SJI 1330~{\AA}, 1400~{\AA} and 2796~{\AA}. SJI observations
have a cadence of about 20 s and a field view of about 40\arcsec
$\times$ 40\arcsec, as shown by the blue box marked in Fig.~\ref{fil}.
The SJI images missed the data at the bottom. The CBP is bright at 7
AIA EUV and 2 SJI FUV bands, and it is very weak at AIA 1600~{\AA},
1700~{\AA} and SJI 2796~{\AA}.

\section{Data Reduction}
\begin{figure}
 \includegraphics[width=\columnwidth]{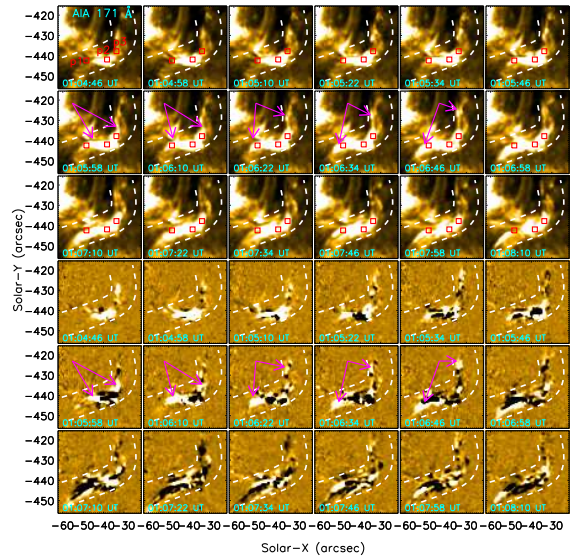}
 \caption{Time sequence of AIA 171~{\AA} images (the top 18 panels)
and their running-difference images (the bottom 18 panels). Each
image has the same size of 40\arcsec $\times$ 40\arcsec. Two dashed
curves outline the curved slit. The three red boxes indicate the
regions (2.4\arcsec $\times$ 2.4\arcsec) used to do the DEM analysis
in Fig.~\ref{dem}. The purple arrows mark a pair of the moving structures (flows).}
 \label{snap}
\end{figure}

As mentioned above, both SDO/AIA and IRIS/SJI have high time
resolution, which gives us an opportunity to study the evolution of
this CBP in detail. Fig.~\ref{snap} shows the time evolution of the
CBP at AIA 171~{\AA}. The top-half of the 18 panels display the
intensity images from 01:04:46~UT to 01:08:10~UT, including the CBP
maximum at $\sim$01:06~UT. There are some bright structures on these
images. These structures are almost simultaneously moving outward
from the CBP core (marked by the red plus in Figs.~\ref{fil} and \ref{cbp}).
For example, the purple arrows mark the two moving structures. They are moving away
separately from the CBP core, one towards northwest, another towards
southeast. Actually, all of the bright structures are moving not
along a straight path, but a curved way. Therefore, this CBP
displays a curved shape in the images with high spatial resolution.
These moving structures start with a bright patch, and their
brightness decreases with the distance from the core, then disappear
far from the core. The other 18 panels of Fig.~\ref{snap} are the
running-difference images at the same time interval. In these
images, one moving structure is divided into one white and one dark
kernels. The white kernel is followed by the dark one, and the
positions of both kernels are changing simultaneously,
as shown by the purple arrows.

Fig.~\ref{snap} just displays the CBP evolution at AIA 171 {\AA}. In
order to display this CBP at other wavelengths, we plot the
time-distance slices along an artificial slit following the curved
shape of this CBP. Firstly, the slit is outlined along the curved
shape with two white dashed lines, as shown in Figs.~\ref{cbp} and
\ref{snap}. Here we use a constant width of $\sim$12\arcsec\ in
order to cover the bulk of the CBP brightness as much as possible
during its lifetime at various wavelengths. Secondly, the
intensities between these two dashed curves are integrated. Thus, we
can obtain the time-distance slices of the intensities at 7 AIA EUV
bands and 2 SJI FUV channels, as shown in Fig.~\ref{slc}. The Y-axis
is the artificial slit from southeast (bottom) to northwest (top).
Thus, the middle of Y-axis, i.e., $\sim$20\arcsec\ is the core
position of this CBP. This CBP begins to become bright at around
01:00~UT, then gradually expands, and significantly enhances almost
simultaneously ($\sim$01:04~UT) at all the 9 wavelengths. The moving
structures from the CBP core are identified as the oblique streaks
on the time-distance slices. They are considered as the flows
\citep[e.g.,][]{Ning14}. Some of them are moving upwards, some
others are moving downwards, as indicated in Fig.~\ref{slc}. In
order to detect these flows, the time derivative is calculated along
the time axis on the time-distance slices. Fig.~\ref{slcd} shows the
derivative time-distance slices at 9 wavelengths. The flows are
still identified as the oblique streaks, which are shown clearly and
each one is labeled by the arrow. In this case, one flow is divided
into two streaks, a white streak followed by a dark one. Similar as
the method used to study the thread in a prominence
\citep{Ning09a,Ning09b}, a pair of white and dark streaks here are
consistent with the movement of a pair of white and dark kernels on
the running-difference images in Fig.~\ref{snap}. For example, a
purple arrow marks one flow moving outward southeast, it corresponds
to the oblique streak labeled by the purple arrow downward in
Fig.~\ref{slcd}, with a speed of about 331~km~s$^{-1}$. Another
purple arrow refers to the flow moving to northwest at the same
time, it corresponds to another oblique streak marked by the purple
arrow upward, with a speed of around 234~km~s$^{-1}$. This pair of
bi-directional flows are detected at all the 9 wavelengths, as
indicated by the bi-directional purple arrows. We also find another
pair of such bi-directional flows which are detected at all the 9
wavelengths, as shown by the pairs of blue arrows in Fig.\ref{slcd}.
Both of these two pairs of bi-directional flows appear at the peak
time of this CBP.

\section{Results}
\subsection{The observational features}
Figs.~\ref{slc} and \ref{slcd} show that the flows move upward
(northwest) and downward (southeast) simultaneously. As marked by
the bi-directional arrows, the flows appear with a pair, and they
start from the CBP core (near about y=20\arcsec). Meanwhile, a pair
of the bi-directional flows have a similar speed. They are
symmetrical on the time-distance slice. There are a number of
bi-directional flows seen at AIA 171~{\AA}, while some of them
display very weak signature at AIA 94~{\AA} and 335~{\AA}. This
indicates that these bi-directional flows tend to be bright at the
upper chromosphere, transition region or the lower corona. The average moving velocity
(speed) of these flows (including the upflows and downflows) is
estimated to about 300~km s$^{-1}$ here, and this value is similar
to that in the outflows detected in the solar flares
\cite[see.,][]{Wang07,Li09,Liu13,Su13,Young13,Ning16}.

Except for these flows are bi-directional, simultaneous, and
symmetrical, they are intermittently and exhibit the periodicity,
whatever upward and downward flows. Fig.~\ref{qpp1} (top panels)
plots the light curves at the position marked by the purple line in
Fig.~\ref{slc}. There are several subpeaks on these light curves at
different wavelengths, and each subpeak indicates a flow. In order
to study these subpeaks' periodicity, the light curve is decomposed
into a slowly varying and a rapidly varying components. The slowly
varying component is the smoothing original flux, as shown by the
blue dashed line overplotted on the light curve. Noting that the
smoothing window is different for the data with different time
resolution \citep[see also.,][]{Li15}. For example, it is 10 points
for AIA data (12~s $\times$ 10) while 6 points for SJI data (20~s
$\times$ 6). The smoothing window has a same interval of 120~s for
both data. The rapidly varying component is the light curve
subtracted by the slowly varying component. Then the wavelet
analysis is applied to detect the periods in the rapidly varying
component. Fig.~\ref{qpp1} presents the results at 9 wavelengths.
Top panels show the light curves (solid) along the purple line in
Fig.~\ref{slc} and the blue dashed profiles are the slowly varying
components. Middle panels display the rapidly varying components
with plus signs. The wavelet spectra are shown at the bottom. A
typical period of around 90~s are detected for all these 9
wavelengths, indicating that this CBP has the quasi-periodic
oscillations, which is similar to the behaviors of quasi-periodic
oscillations in the solar flares \citep{Li15,Ning16}.

Using the same method, Fig.~\ref{qpp2} displays the analysis results
of the light curves at the position indicated by the turquoise line
in Fig.~\ref{slc}. And the same period as about 90~s is detected
again. As marked by the purple and turquoise lines, they are
symmetrically located at two sides of the CBP core. The different
peaks in Fig.~\ref{qpp1} represent the various flows towards the
northwest, while those in Fig.~\ref{qpp2} are the flows towards the
southeast. The same periods detected in them not only indicate the
periodicity, but also further confirm the symmetrical property of
the bi-directional flows.

\subsection{The magnetic field}
Fig.~\ref{hmi} shows the LOS magnetograms from HMI at four different
times, which cover the entire lifetime of the CBP. Except for the
positive and negative fields within the big region, which are stable
before and after the CBP, there is a flashed positive field with a
small region like 4.8\arcsec$\times$7.2\arcsec, as marked by the
blue box. The contours are set at 20~G. Fig.~\ref{light} (bottom)
shows the time evolution of this positive filed strength (black profiles). It seems
to enhance at about 01:03~UT, and reach its maximum at about
01:06:30~UT, while almost disappear at around 01:08~UT. It has a
lifetime of $\sim$5 minutes, which covers the CBP lifetime very
well. The CBP takes place at the position between this flashed
positive field and negative field. The plus marks the core where the
CBP starts. Meanwhile, this positive filed moves towards the
negative filed during the CBP lifetime.

Fig.~\ref{slice} (bottom) shows the time-distance slice along a
directional slit, as indicated by two white solid lines in
Fig.~\ref{hmi}. It is from 00:55~UT to 01:18~UT, which covers the
CBP whole lifetime. Except for this positive field decreases its
strength (as indicated by the contours), it moves close towards the
negative field with a stable speed of about 1.2~km~s$^{-1}$ (blue
arrow). Actually, the negative filed also moves close to the
positive filed at a speed of around 2.6~km~s$^{-1}$ (red arrow),
then stops after the CBP maximum. These facts suggest the magnetic
cancelation occurrence at the position of the CBP core. Our findings
are agreement very well with previous results about the relationship
between the CBPs and magnetic fields
\citep[e.g.,][]{Golub76,Harvey85,Webb93,Li12,Zhang12,Chen15}. All
these studies found that the CBPs were related to the small-scale
magnetic fields, especially the magnetic dipoles. Which means that
the CBPs are originating from the small-scale magnetic
reconnections.

\subsection{The DEM analysis}
In order to investigative the temperature and emission measure (EM)
of the CBP, we perform the DEM analysis using the six AIA EUV bands,
such as 94~{\AA}, 131~{\AA}, 171~{\AA}, 193~{\AA}, 211~{\AA}, and
335~{\AA} \citep{Cheng12}. Firstly, we have to measure their fluxes
at the small regions, such as p1, p2 and p3 labeled in
Fig.~\ref{snap}. The region of p2 is located at the CBP core, while
p1 and p3 regions are outside of the CBP. Then the function of the
DEM dependence on the temperature is fitting from the observations.
Fig.~\ref{dem} shows the DEM curves of the p2 region at four times.
The best-fit DEM solution to the observed fluxes are indicated with
the black solid profiles. And the 100 Monte Carlo (MC) realizations
of the data are computed to estimate the DEM uncertainties, as
indicated by the color rectangles. The yellow rectangles represent
the region surrounding the best-fit solution that contains 50\% of
the MC solutions, and two turquoise and one yellow rectangles are
consisted of the regions that covers 80\% of the MC solutions, while
all of the colored rectangles include the regions which contain 95\%
of the MC solutions. Therefore, the top and bottom ends of the
colored rectangles can be regarded as the uncertainties of the DEM
analysis. Given the uncertainties of the DEM analysis, the average
temperature ($T$) and emission measure (EM) at each time are
calculated over the temperature range from log $T$ = 5.7 to log $T$ =
7.0.

Fig.~\ref{em_te} shows the time evolution of the average temperature
(top) and EM (bottom) at three regions of p1, p2 and p3. The CBP
core (p2) increases its temperature firstly (at $\sim$01:03~UT) and
rapidly, while the p1 and p3 regions increase their temperature
about 2 minutes later, at around 01:05~UT. This is consistent with
the moving structures from the CBP core towards two sides. Similarly
for the EM, p1 and p3 regions reach their peak values after the CBP
core (p2), and the peak EM at the CBP core is almost at the same
time as the AIA 171 {\AA} maximum brightness (blue profile). The
maximum values of both temperature and EM are at the CBP core
region, and their values are about 3.148~MK and
9.79$\times$10$^{27}$~cm$^{-5}$, which are in agreement with
previous findings in the CBPs
\citep[e.g.,][]{Nolte79,Tian08,Kariyappa11,Hong14,Zhang14b,Zhang16}.
This DEM result is also consistent with the CBP behaviors at the
various solar atmospheres. That is to say, the CBP can radiate
stronger emissions at AIA 171~{\AA}, 193~{\AA}, 211~{\AA},
131~{\AA}, 304~{\AA} and SJI 1330~{\AA}, 1400~{\AA}, all these bands
can detect the emissions from upper chromosphere, transition region
or corona (quiet and active) at solar surface, which correspond the
temperature of around 0.1$-$1.0 MK. But it radiates the weak
emissions at AIA 94 and 335~{\AA}, because these two bands mainly
detect the emissions from the flaring corona which have a higher
temperature. And it has nearly no radiations at AIA 1600~{\AA},
1700~{\AA} and SJI 2796~{\AA}, whose emissions are mostly from the
photosphere or lower chromosphere, and have a lower temperature
(below 10$^4$ K). All these findings indicate that the CBP has a
multi-temperature structures (see also., Fig.~\ref{dem}).

\section{Conclusions and Discussions}
Based on the observations from {\it IRIS} and {\it SDO}, we explore
the bi-directional moving structures in a CBP at multi-bands, i.e.,
7 AIA EUV and 2 SJI FUV bands. This CBP takes place at NOAA 12384
with a curved shape. AIA and SJI observations show that the CBP is
brightening at AIA EUV and SJI FUV wavelengths. Meanwhile, the
observations with high temporal resolution show that there are many
bright structures moving intermittently from the CBP core outward
two sides. From the time-distance slices along a curved slit, these
moving structure are identified as the bi-directional flows toward
northwest and southeast simultaneously. They display the similar
shape and velocity, but opposite directions, and their average moving
speed is around 300~km~s$^{-1}$. They also show a typical periodicity with
a quasi-period of $\sim$90~s.

\begin{figure}
 \includegraphics[width=\columnwidth]{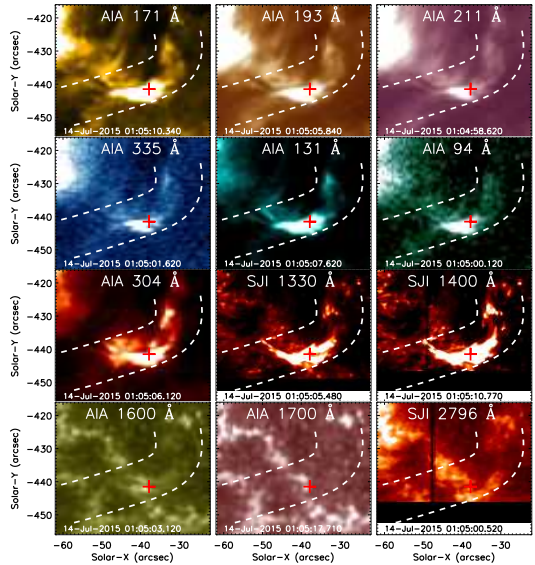}
 \caption{The CBP at SDO/AIA and IRIS/SJI images. Two dashed curves
outline the curved slit where the CBP takes place. The red plus (`+')
marks the CBP core.}
 \label{cbp}
\end{figure}

\begin{figure}
 \includegraphics[width=\columnwidth]{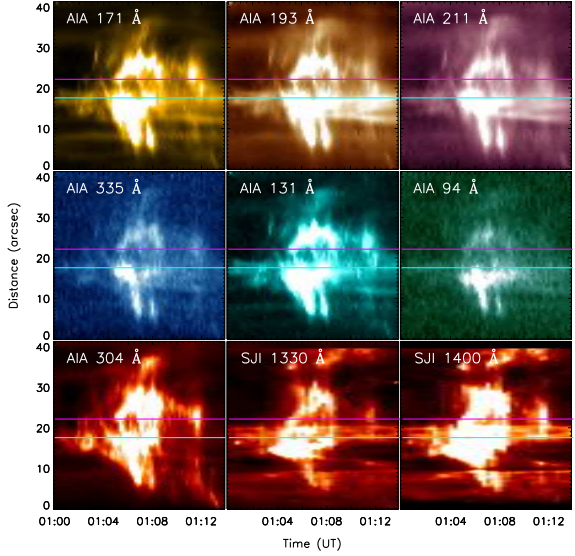}
 \caption{Time-distance slices at 7 AIA EUV bands and 2 SJI
wavelengths along the curved slit in Fig.~\ref{cbp}. The purple and
turquoise lines indicate the two symmetrical positions whose light
curves are studied in Fig.~\ref{qpp1} and Fig.~\ref{qpp2}.}
 \label{slc}
\end{figure}

\begin{figure}
 \includegraphics[width=\columnwidth]{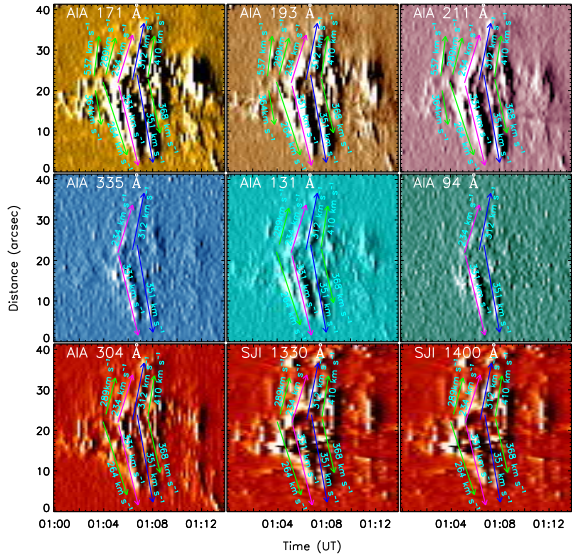}
 \caption{Derivative (gradient along the time axis) of the
time-distance slice in Fig.~\ref{slc}. The arrows mark the various
pairs of the moving structures (flows).}
 \label{slcd}
\end{figure}

\begin{figure}
 \includegraphics[width=\columnwidth]{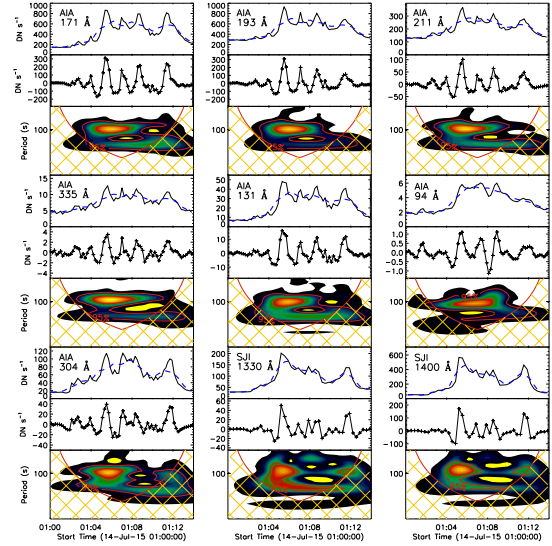}
 \caption{Wavelet analysis of the light curves at 9 wavelengths.
Top: light curves (solid) with the slowly varying components
(dashed) at the position marked by the purple line in
Fig.~\ref{slc}. Middle: the rapidly varying components (plus).
Bottom: wavelet spectra.}
 \label{qpp1}
\end{figure}

\begin{figure}
 \includegraphics[width=\columnwidth]{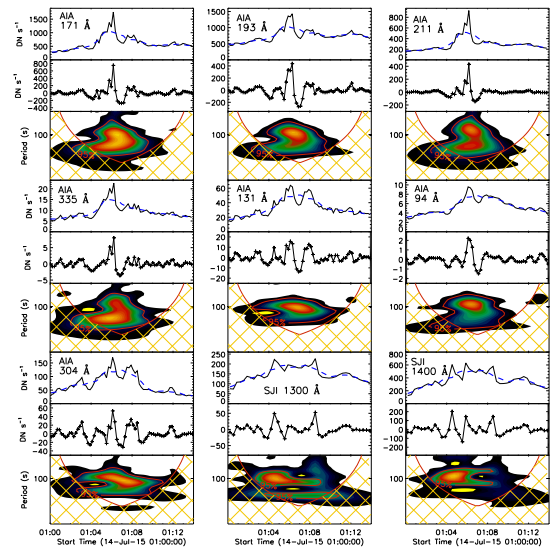}
 \caption{Same as Fig.~\ref{qpp1}, but the analysis is performed at
the position marked by the turquoise lines in Fig.~\ref{slc}.}
 \label{qpp2}
\end{figure}

\begin{figure}
 \includegraphics[width=\columnwidth]{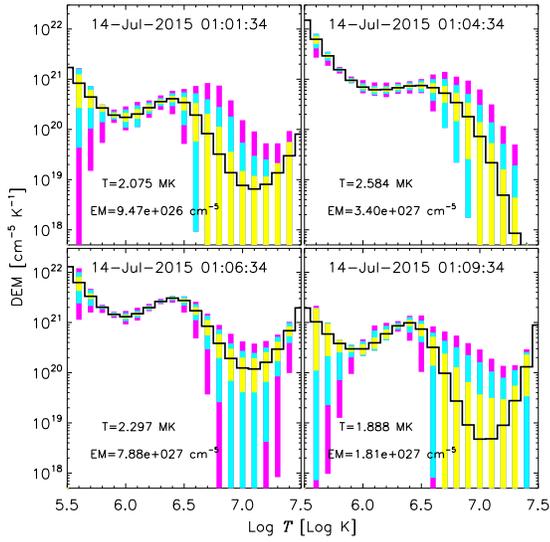}
 \caption{The DEM profiles of the CBP core (p2) at four times.
The black solid profiles give the best-fitted DEM curves from the
observations. The yellow rectangles represent the regions that
contains 50\% of the MC solutions. The turquoise rectangles, above
and below the yellow rectangles, and the yellow rectangles compose
the regions that cover 80\% of the MC solutions. All of the colored
rectangles form the regions containing 95\% of the MC solutions. The
average temperature and EM are given.}
 \label{dem}
\end{figure}

\begin{figure}
 \includegraphics[width=\columnwidth]{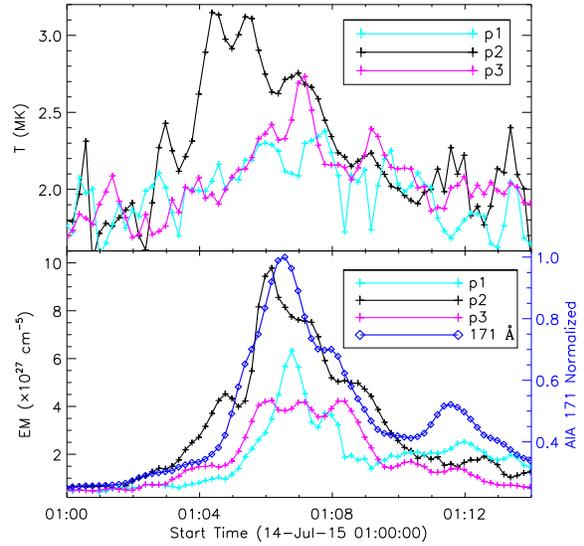}
 \caption{The light curves of the temperature (top) and EM (bottom)
from the DEM analysis at three regions of p1, p2 and p3. The 171
normalized flux (blue) is given.}
 \label{em_te}
\end{figure}

In our results, at least two pairs of bi-directional flows are
detected simultaneously at all the 9 wavelengths from AIA and SJI
observations, as indicated by the purple and blue arrows in
Fig.~\ref{slcd}. This finding fills the gap that the bi-directional
flows are not observed simultaneously at all 9 AIA wavelengths
\citep{Ning14}. Here, we use the observations from SJI FUV images
instead of AIA UV images. This is because the CBP has very weak
radiations at AIA UV bands (see., Fig.~\ref{cbp}). Thus, the
bi-directional flows induced by the instrument self-effects can also
be ruled out, since observations are from two different instruments
which onboard {\it SDO} and {\it IRIS}, respectively. And the
quasi-period of $\sim$90~s are also detected simultaneously at all
these 9 wavelengths at the symmetrical positions from the CBP core
(see., Fig.~\ref{qpp1} and \ref{qpp2}), suggesting the
quasi-periodic oscillations in this CBP. This is similar as previous
findings \citep[e.g.,][]{Nolte79,Tian08b,Ning14,Zhang14a,Samanta15},
and maybe related with the reciprocatory magnetic reconnection
\citep[e.g.,][]{Priest94,Zhang14,Chen15}. The period could be
modulated by certain MHD wave \citep{Chen06,Banerjee16}. The
simultaneous, symmetry and quasi-periodic bi-directional flows in a
CBP can be detected at all the 9 wavelengths, suggesting that the
flows could be existed at various solar atmosphere layers. For
example, the chromosphere and transition region (AIA 131 and
304~{\AA}, SJI 1330 and 1400~{\AA}), the quiet corona (AIA
171~{\AA}), the active-region corona (AIA 193, 211 and 335~{\AA}),
and even the flaring corona (AIA 94, 131, and 193~{\AA})
\citep{Lemen12,Dep14}. There are much less and weaker flows detected
at AIA 94 and 335~{\AA} (Fig.~\ref{slc} and \ref{slcd}), indicating
that the flows mainly behaving at the transition region and corona,
further suggesting that the temperature of the CBP is around 1.0 MK,
which is similar to the DEM analysis results (see., Fig.~\ref{dem}
and \ref{em_te}). This is also consistent with previous findings
about the study of CBP temperature
\citep{Nolte79,Tian08,Kariyappa11,Zhang14b}.

The average velocity of these bi-directional flows is around 300 km
s$^{-1}$ (see., Fig.~\ref{slcd}), this value has the same order of
the outflows which detected in the solar flares
\citep[e.g.,][]{Wang07,Li09,Liu13,Su13,Young13,Ning16}. On the other
hand, statistical results find that the CBPs and solar flares
display the similar power-law behaviors, and the power-law index is
closing to 2.0 \citep[e.g.,][]{Dennis85,Crosby93,Aschwanden98,Li13}.
All these similar findings between the CBPs and solar flares suggest
that there is no fundamental difference to generate them. Researches
further indicate that they are strongly related to the magnetic
structures (small or large scales) at the solar surface. In other
words, the mechanisms of the magnetic features which related to the
solar eruptions (CBPs and solar flares) on solar disk are scale-free
\citep[see also.,][]{Parnell09}.

\begin{figure}
 \includegraphics[width=\columnwidth]{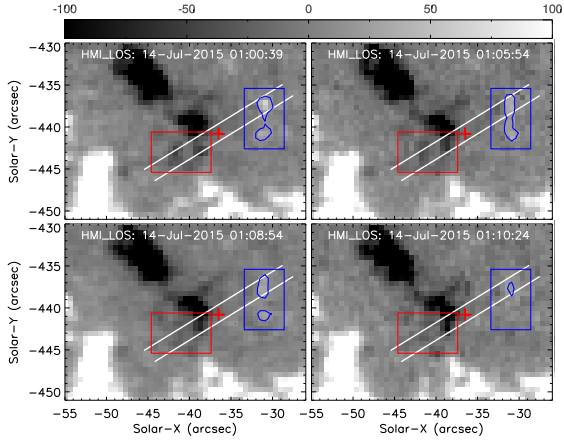}
 \caption{HMI magnetograms at four times. The
contours represent the positive fields at the levels of 20~G. Each
image has the same size of 29\arcsec $\times$ 21\arcsec. Two solid
lines outline the slit used in Fig.~\ref{slice}. The plus (`+')
indicates the CBP core. The rectangles mark the regions used to
calculate the magnetic strength of positive (blue) and negative
(red) fields.} \label{hmi}
\end{figure}

\begin{figure}
 \includegraphics[width=\columnwidth]{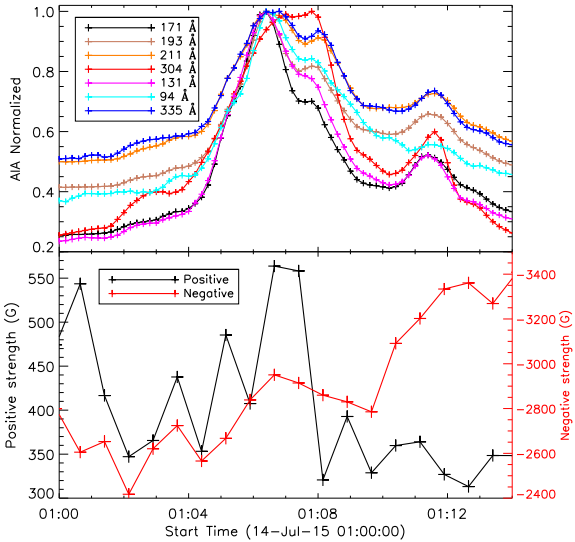}
 \caption{Top: the normalized light curves along the curved slice at 7 AIA EUV
channels. Bottom: HMI magnetic filed strength at the rectangle
regions in Fig.~\ref{hmi}.}
 \label{light}
\end{figure}

\begin{figure}
 \includegraphics[width=\columnwidth]{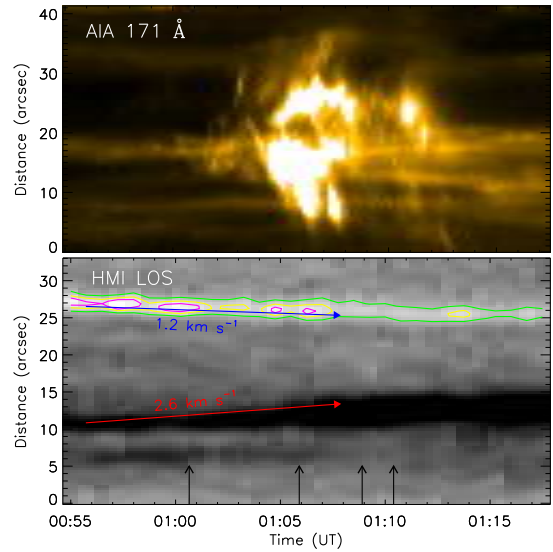}
 \caption{Top: Time-distance slice at AIA 171 {\AA} along the curved
slit in Fig.~\ref{cbp}. Bottom: Time-distance slice along the tilt
slit in Fig.~\ref{hmi}. The contours represent the positive fields
at the level of 55~G (green), 95~G (yellow) and 120~G (purple),
respectively. The color arrows mark the movement of the positive
(blue) and negative (red) magnetic fields, respectively. HMI images
at four times marked by the black arrows are given in
Fig.~\ref{hmi}}
 \label{slice}
\end{figure}

HMI observations show that a flashed positive field emerges near the
CBP core (Fig.~\ref{hmi}). And it has a short lifetime, same as the
CBP interval. Meanwhile, it moves close towards the nearby negative
field, and this CBP takes place between them (Fig.~\ref{light} and
\ref{slice}). Furthermore, the CBP core increases its temperature
and emission measure earlier than the outside regions
(Fig.~\ref{em_te}). All these facts support that the magnetic
reconnection model can explain this CBP and the bi-directional
flows. It is well known that the magnetic reconnection is the basic
process which the magnetic energy convert into kinetic energy at the
solar atmosphere \citep{Innes97}. And this magnetic reconnection
model has been used to explain the solar eruptions in various
scales, such as the solar flares and CMEs in large scale
\citep{Su13,Liu13,Li15,Ning16}, CBPs and transition explosive evens
in small scale \citep{Perez99,Zhang12,Innes13,Ning14,Innes15}.
However, the direct observation evidences which support the magnetic
reconnection are rare, because of the lower spatial and temporal
resolution observations. \cite{Innes97} had indicated that the
bi-directional outflows in the transition explosive events can be as
the direct observational evidence of the magnetic reconnection. This
is consistent with the MHD results, which suggest that the
bi-directional outflows in the explosive evens are from magnetic
reconnection regions \citep{Innes99,Sarro99,Roussev01a,Roussev01b}.
Recently, \cite{Ning14} detected the similar bi-directional outflows
in a CBP, and considered them as the direct observational evidence
of the magnetic reconnection. Therefore, the bi-directional flows in
this CBP could be the bi-directional outflows, and they can be as
the direct observation evidence of the magnetic reconnection at
solar surface. Considering the radiations from this CBP are mainly
from the transition region and lower corona, together with the
statistical results that there are hundreds of CBPs emergence at
solar surface at any given time \citep{Zhang01}, they could be
contributed to the coronal heating \citep{Priest94}.

In this paper, we report the bi-directional outflows and their
quasi-periodicity in a CBP observed at by {\it SDO}/AIA and {\it
IRIS}/SJI. They could be explained by the reconnection model. This
is similar as the bi-directional outflows in the explosive events
which also interpreted as the reconnection jets
\citep[e.g.,][]{Innes97,Innes99,Sarro99,Perez99,Roussev01a,Roussev01b,Innes13,Innes15}.
As indicated by \cite{Chae98}, the explosive events may be the
manifestation of hot plasma materials flowing out of the transition
region. Therefore, the outflows in this CBP could represent the hot
plasmas flowing out of the solar atmospheres, including the upper
chromosphere, transition region and lower corona. In other words,
the bi-directional outflows in explosive events and CBPs may have
the similar physical mechanism, but they display the different
observational features at the spectral and imaging observations
\citep[see also.,][]{Chae00}. In future, the joint observations from
the {\it IRIS} spectra and {\it SDO} images could be a good topic to
study their originations. The quasi-period of about 90~s may be
modulated by some MHD waves \citep{Chen06,Banerjee16}. However, it
is hard to determine the MHD waves which could play the key effects
here, because of the observation limits. Therefore, future work will
also focus on the analysis of the wave-modulation causing the
periodicity in CBPs or explosive events.

\acknowledgments We would like to thank the anonymous referee for
his/her valuable comments to improve the manuscript. The data used
in this paper are mainly from {\it IRIS} and {\it SDO}. {\it IRIS}
is a NASA small explorer mission developed and operated by LMSAL
with mission operations executed at NASA Ames Research center and
major contributions to downlink communications funded by the
Norwegian Space Center (NSC, Norway) through an ESA PRODEX contract.
{\it SDO} is the first mission to be launched under NASA's Living
With a Star (LWS) program. This study is supported by NSFC under
grants 11333009, 11303101, 11473071, 11573072, 973 program
(2014CB744200) and Laboratory NO. 2010DP173032. This work is also
supported by the Youth Fund of Jiangsu No. BK20141043, BK20161095,
and one hundred talent program of Chinese Academy of Sciences.

\clearpage

\end{document}